\documentclass{jpsj-suppl}
\usepackage{txfonts} 
\usepackage{amsmath,amsfonts,amssymb,bm}
\usepackage{dsfont}
\usepackage{graphicx}
\usepackage{color}
\usepackage{soul}

\usepackage[mathscr,scaled=1.15]{urwchancal}
\DeclareFontFamily{OT1}{pzc}{}
\DeclareFontShape{OT1}{pzc}{m}{it}%
{<-> s * [1.15] pzcmi7t}{}
\DeclareMathAlphabet{\mathpzc}{OT1}{pzc}{m}{it}

\definecolor{purple}{rgb}{0.5,0,0.5}
\definecolor{blue}{rgb}{0.0,0,0.9}

\newcommand{\lsim}{\mathrel{\rlap{\lower4pt\hbox{\hskip0pt$\sim$}}
\raise1pt\hbox{$<$}}}           
\newcommand{\gsim}{\mathrel{\rlap{\lower4pt\hbox{\hskip0pt$\sim$}}
\raise1pt\hbox{$>$}}}           

\title{Running Masses in the Nucleon and its Resonances}

\author{Craig D.~\textsc{Roberts}}

\inst{Physics Division, Argonne National Laboratory, Argonne, Illinois 60439, USA}

\email{cdroberts@anl.gov}

\recdate{15 September 2015}

\abst{An overarching scientific challenge for the coming decade is to discover the meaning of confinement, its relationship to dynamical chiral symmetry breaking (DCSB) -- the origin of visible mass -- and the connection between them.  In progressing toward meeting this challenge, significant progress has been made using continuum methods in QCD.  For example, a novel understanding of gluon and quark confinement and its consequences has begun to emerge from quantum field theory; a clear picture is being drawn of how hadron masses emerge dynamically in a universe with light quarks; and ground-state hadron wave functions with a direct connection to QCD are becoming available, which reveal that quark-quark correlations are crucial in hadron structure.  There is growing experimental support for this body of predictions in both elastic and nucleon-to-resonance-transition form factors.}

\kword{confinement, diquark clusters, dynamical chiral symmetry breaking (DCSB), nucleon elastic and transition form factors}

\begin{document}
\maketitle

\section{Confinement and DCSB}
\label{secConfinement}
Most hadron physicists have a notional understanding of confinement; but, in order to consider the concept in depth, it is crucial to have a concrete definition.  That problem is canvassed in Sec.\,2.2 of Ref.\cite{Cloet:2013jya}, which explains that the potential between infinitely-heavy quarks measured in simulations of quenched lQCD -- the so-called static potential -- is \emph{irrelevant} to the question of confinement in our Universe, in which light quarks are ubiquitous and the pion is unnaturally light.  This is because light-particle creation and annihilation effects are essentially nonperturbative and so it is impossible in principle to compute a quantum mechanical potential between two light quarks \cite{Bali:2005fuS, Prkacin:2005dc, Chang:2009ae}.  There is no flux tube in a Universe with light quarks, the flux tube is not a valid paradigm for confinement, and hence it is meaningless to speak of linear potentials and Regge trajectories \cite{Tang:2000tb, Masjuan:2012gc}.

DCSB is critical here.  It ensures the existence of nearly-massless pseudo-Goldstone modes (pions), each constituted from a valence-quark and -antiquark whose individual current-quark masses are $<1$\% of the proton mass \cite{Qin:2014vya}.  These modes ensure that no flux tube between a static colour source and sink can have a measurable existence.  To see this, consider such a tube being stretched between a source and sink.  The potential energy within the tube may increase only until it reaches that required to produce a particle-antiparticle pair of the theory's pseudo-Goldstone modes.  Simulations of lQCD show \cite{Bali:2005fuS, Prkacin:2005dc} that the flux tube then disappears instantaneously along its entire length, leaving two isolated colour-singlet systems.  The length-scale associated with this effect in QCD is $r_{\not\sigma} \simeq (1/3)\,$fm.  Hence if any such string formed, it would dissolve well within a hadron's interior.

An alternative perspective associates confinement with dynamically-driven changes in the analytic structure of QCD's propagators and vertices.  In this realisation, \emph{confinement is a dynamical process}, whose expression cannot be understood using models that violate Poincar\'e-covariance or any other symmetry that is crucial to the observable features of hadrons.  Modern theory predicts that both gluons and quarks acquire mass distributions, which are large at infrared momenta \cite{Bhagwat:2003vw, Bhagwat:2006tu, Bowman:2005vx, Binosi:2014aea}.  These running masses lead to the emergence of a length-scale $\varsigma \approx 0.5\,$fm, whose existence and magnitude are evident in all studies of dressed-gluon and -quark propagators and which characterises a dramatic change in their analytic structure.  In models based on such features \cite{Stingl:1994nk}, once a gluon or quark is created, it begins to propagate; but after each ``step'' of length $\varsigma$, on average, an interaction occurs so that the parton loses its identity, sharing it with others.  Finally a cloud of partons is produced, which coalesces into colour-singlet final states.  Such pictures of parton propagation, hadronisation and confinement can be tested in experiments at modern and planned facilities, \emph{e.g}.\ via measurements that chart parton distribution amplitudes and functions of mesons, and the nucleon and its excited states.

Whilst the nature and realisation of confinement in empirical QCD is still being explored, DCSB; namely, the generation of \emph{mass} \emph{from} \emph{nothing}, is a theoretically-established feature of QCD.  It is ``dynamical,'' as distinct from spontaneous, because nothing is added to QCD in order to effect this remarkable outcome and there is no simple change of variables in the QCD action that will make it apparent.  Instead, through the act of quantising the classical chromodynamics of massless gluons and quarks, a large mass-scale is generated.  DCSB is the most important mass generating mechanism for visible matter in the Universe, being responsible for approximately $98$\% of the proton's mass.

\begin{figure}[t]
\begin{minipage}[t]{\textwidth}
\begin{minipage}{0.5\textwidth}
\centerline{\includegraphics[width=0.9\textwidth]{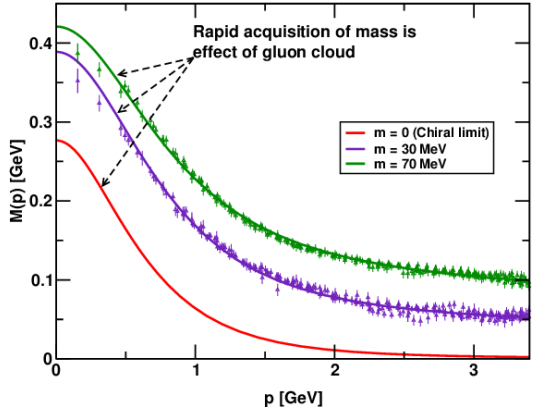}}
\end{minipage}
\begin{minipage}{0.5\textwidth}{\small
\caption{\label{gluoncloud} \small
Dressed-quark mass function, $M(p)$ in Eq.\,(\ref{SgeneralN}): \emph{solid curves} -- DSE results, explained in Refs.\,\protect\cite{Bhagwat:2003vw,Bhagwat:2006tu}, \emph{points} -- numerical simulations of lattice-regularised QCD \protect\cite{Bowman:2005vx}.  (\emph{N.B}.\ $m=70\,$MeV is the uppermost curve and current-quark mass decreases from top to bottom.)  The current-quark of perturbative QCD evolves into a constituent-quark as its momentum becomes smaller.  The constituent-quark mass arises from a cloud of low-momentum gluons attaching themselves to the current-quark.  This is DCSB: an essentially nonperturbative effect that generates a quark \emph{mass} \emph{from nothing}; namely, it occurs even in the chiral limit.
}}
\end{minipage}
\end{minipage}
\end{figure}

A fundamental expression of DCSB is the behaviour of the quark mass-function, $M(p)$, which is a basic element in the dressed-quark propagator:
\begin{equation}
\label{SgeneralN}
S(p) = 
1/[i\gamma\cdot p A(p^2) + B(p^2)] = Z(p^2)/[i\gamma\cdot p + M(p^2)]\,,
\end{equation}
and may be obtained as a solution to QCD's most basic fermion gap equation, \emph{i.e}. the Dyson-Schwinger equation (DSE) for the dressed-quark propagator \cite{Cloet:2013jya}.  The nontrivial character of the mass function in Fig.\,\ref{gluoncloud} arises primarily because a dense cloud of gluons comes to clothe a low-momentum quark.  It explains how an almost-massless parton-like quark at high energies transforms, at low energies, into a constituent-like quark with an effective ``spectrum mass'' $M_D \sim 350\,$MeV.  

\section{Gluon Cannibalism}
\label{secCannibals}
The propagation of gluons, too, is described by a gap equation \cite{Aguilar:2009nf}; and its solution shows that gluons are cannibals: they are a particle species whose members become massive by eating each other!  The gluon mass function, $m_g(k^2)$, is monotonically decreasing with increasing $k^2$, with $m_g(0) \approx 0.5$GeV \cite{Binosi:2014aea}.
The mass term appears in the transverse part of the gluon propagator, hence gauge-invariance is not tampered with; and the mass function falls as $1/k^2$ for $k^2\gg m_g(0)$ (up to logarithmic corrections), so the gluon mass is invisible in perturbative applications of QCD. 

Gluon cannibalism presents a new physics frontier within the Standard Model. Asymptotic freedom means that the ultraviolet behaviour of QCD is controllable.  At the other extreme, dynamically generated masses for gluons and quarks entail that QCD creates its own infrared cutoffs.  Together, these effects eliminate both the infrared and ultraviolet problems that typically plague quantum field theories and thereby make reasonable the hope that QCD is nonperturbatively well defined.

The dynamical generation of gluon and quark masses provides a basis for understanding the notion of a maximum wavelength for partons in QCD \cite{Brodsky:2008be}.  Given the magnitudes of these mass-scales, it is apparent that field modes with wavelengths $\lambda > \varsigma \approx 2/m_g(0) \approx 0.5\,$fm decouple from the dynamics: they are screened in the sense described in Sec.\,\ref{secConfinement}.  This is just one consequence of a dynamically generated gluon mass-scale.  There are many more. \emph{e.g}.\ 
it is plausible to conjecture that dynamical generation of an infrared gluon mass-scale leads to saturation of the gluon parton distribution function at small Bjorken-$x$ within hadrons. The possible emergence of this phenomenon stirs great scientific interest and curiosity and it is a key motivation in plans to construct an EIC \cite{Accardi:2012qutS}.

\begin{figure}[t]
\begin{minipage}[t]{\textwidth}
\begin{minipage}{0.48\textwidth}
\centerline{\includegraphics[clip,width=0.9\textwidth]{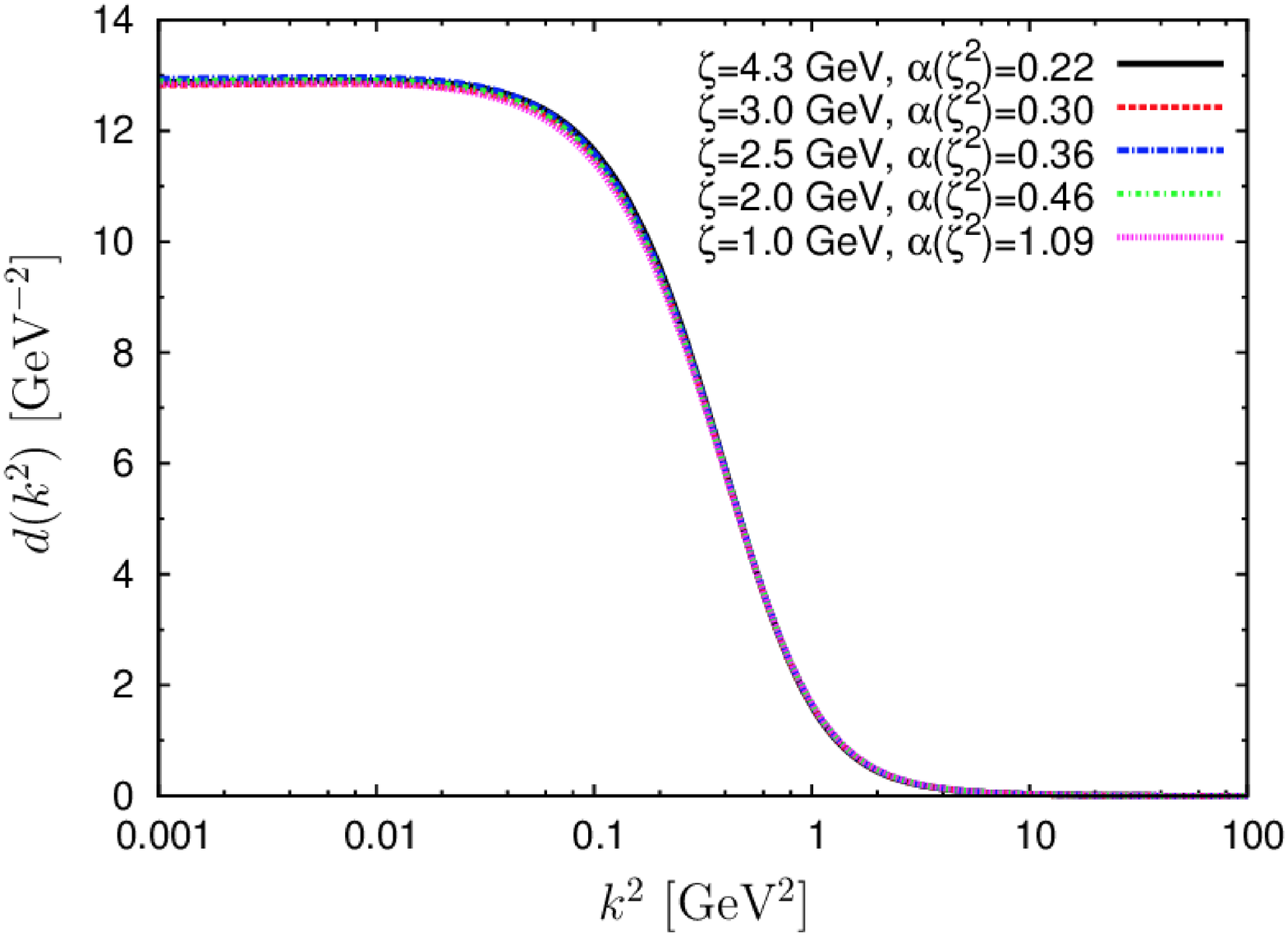}}
\end{minipage}
\begin{minipage}{0.02\textwidth}
\hspace*{-0.2em}\mbox{\LARGE \textbf{$\Rightarrow$}}
\end{minipage}
\begin{minipage}{0.48\textwidth}
\centerline{\includegraphics[clip,width=0.9\textwidth]{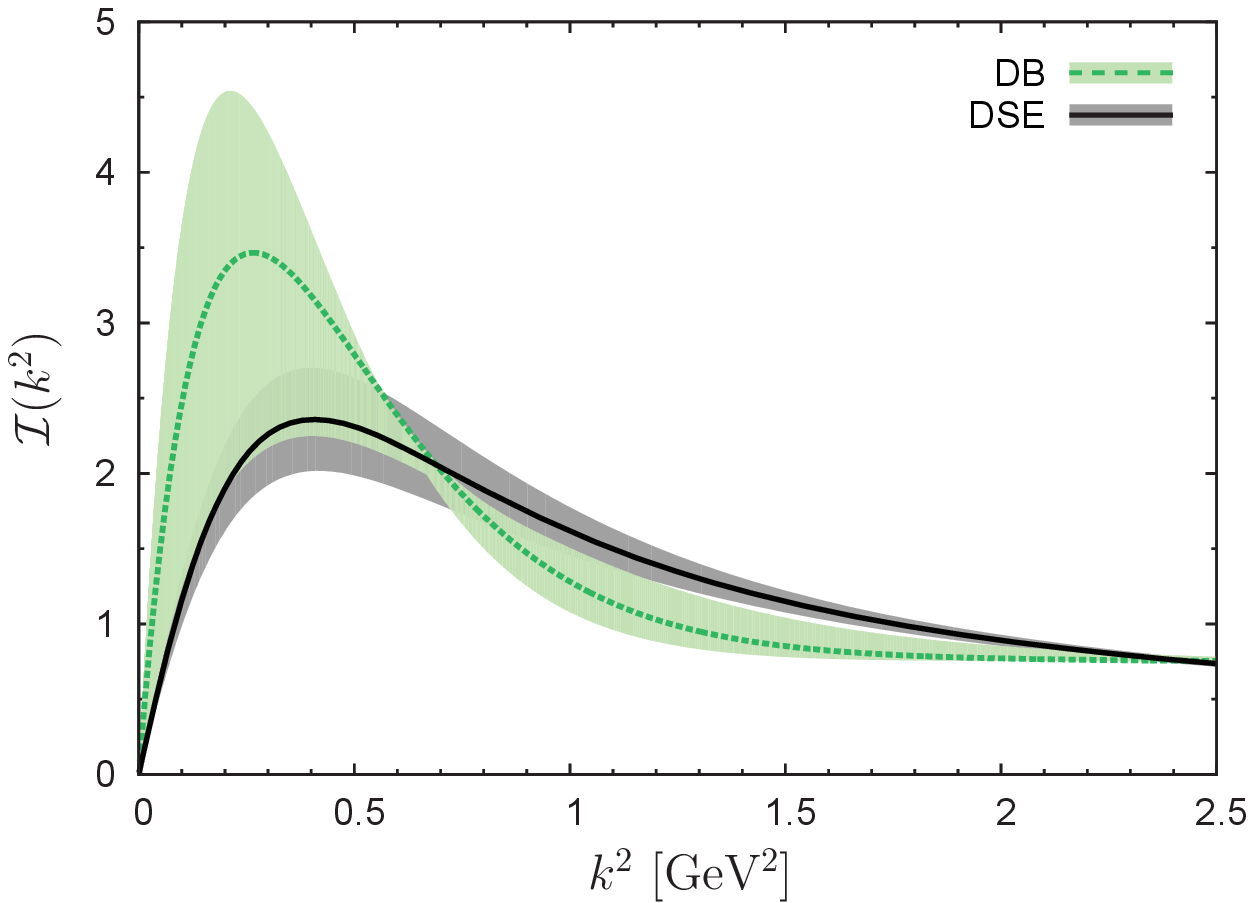}}
\end{minipage}
\end{minipage}
\caption{\label{figInteraction} \small
\emph{Left} -- RGI running interaction computed via a combination of DSE- and lattice-QCD analyses \cite{Aguilar:2009nf}.  The function obtained with five different values of the renormalisation point is depicted in order to highlight that the result is RGI.  The interaction is characterized by $\alpha_s(0) \approx 0.9\, \pi$ and the gluon mass-scale $m_g(0) \approx 0.5$GeV.
\emph{Right} -- Comparison between top-down results for the gauge-sector interaction (derived from the left-panel) with those obtained using the bottom-up approach based on hadron physics observables.  \underline{Solid curve} within \emph{grey band} -- top-down result for the RGI running interaction; and \underline{dashed curve} within \emph{pale-green band} -- advanced bottom-up result obtained using the most sophisticated truncation of the matter sector DSEs - the DSE-DB kernel.  The bands denote the domain of uncertainty in the determinations of the interaction.
}
\end{figure}

\section{Continuum-QCD and \emph{ab initio} predictions of hadron observables}
\label{secAbInitio}
Within hadron physics there are two methods for determining the mo\-men\-tum-dependence of the interaction between quarks: the top-down approach, which works toward an \textit{ab initio} computation of the interaction via analysis of the gauge-sector gap equations; and the bottom-up scheme, which infers the interaction by fitting data within a well-defined truncation of those equations in the matter sector that are relevant to bound-state properties.  These two approaches have recently been united by a demonstration that the renormalisation-group-invariant (RGI) running-interaction predicted by contemporary analyses of QCD's gauge sector coincides with that required in order to describe ground-state hadron observables using a nonperturbative truncation of QCD's DSEs in the matter sector \cite{Binosi:2014aea}.

The unification is illustrated in Fig.\,\ref{figInteraction}: the interaction derived from QCD's gauge sector is in near precise agreement with that required for a veracious description of hadron properties using the most sophisticated matter-sector gap and Bethe-Salpeter kernels available today.  This is a remarkable result, given that there had previously been no serious attempt at communication between practitioners from the top-down and bottom-up hemispheres of continuum-QCD.  It bridges a gap that had lain between nonperturbative continuum-QCD and the \emph{ab initio} prediction of bound-state properties.

It should be noted that if the realistic interaction depicted in Fig.\,\ref{figInteraction} were employed as the seed for a RL-truncation study, it would fail completely because, \emph{inter alia}, DCSB would be absent.  We now know that a veracious description of DCSB and hence hadron properties in QCD requires a dressed-quark-gluon vertex.  Constraining its form is a topic of great contemporary interest; and in this connection it cannot be emphasised too strongly that little of value today will be produced by any attempt at a term-by-term diagrammatic construction of this vertex.

\section{Enigma of mass}
\label{secEnigma}
As noted in Sec.\,\ref{secConfinement}, the pion is Nature's lightest hadron.  In fact, it is peculiarly light, with a mass just one-fifth of that which quantum mechanics would lead one to expect.  This remarkable feature has its origin in DCSB.  The pion's structure is described by a Bethe-Salpeter amplitude:
$\Gamma_{\pi}(k;P) = \gamma_5 [
i E_{\pi}(k;P) + \gamma\cdot P F_{\pi}(k;P)  + \gamma\cdot k \, G_{\pi}(k;P) - \sigma_{\mu\nu} k_\mu P_\nu H_{\pi}(k;P)],$
(here $k$ is the relative momentum between the  valence-quark and -antiquark constituents, and $P$ is their total momentum).
%
In QCD if, and only if, chiral symmetry is dynamically broken, then in the chiral limit \cite{Qin:2014vya}:
\begin{equation}
\label{gtrE}
f_\pi E_\pi(k;0) = B(k^2)\,,
\end{equation}
where $f_\pi$ is the pion's leptonic decay constant, a directly measurable quantity that connects the strong and weak interactions, and the rhs is a scalar function in the dressed-quark propagator, Eq.\,\eqref{SgeneralN}.  This identity is miraculous.  It means that the two-body problem is solved, almost completely once the solution to the one body problem is known.  Eq.\,\eqref{gtrE} is a quark-level Goldberger-Treiman relation.  It is also the most basic expression of Goldstone's theorem in QCD, \emph{viz}.\\[-2ex]

\centerline{\parbox{0.90\textwidth}{\flushleft \emph{Goldstone's theorem is fundamentally an expression of equivalence between the one-body problem and the two-body problem in QCD's colour-singlet pseudoscalar channel.}}}

\medskip

\hspace*{-\parindent}Eq.\,\eqref{gtrE} emphasises that Goldstone's theorem has a pointwise expression in QCD; and, furthermore, that pion properties are an almost direct measure of the mass function depicted in Fig.\,\ref{gluoncloud}.  Thus, enigmatically, properties of the (nearly-)massless pion are the cleanest expression of the mechanism that is responsible for almost all the visible mass in the Universe.  
This provides strong motivation for pion form factor and distribution function measurements at JLab\,12 \cite{E1206101, E1207105, Keppel:2015}.

\section{Structure of Baryons}
It is crucial to address the three valence-quark bound-state problem in QCD with the same level of sophistication that is now available for mesons, with the goal being to correlate the properties of meson and baryon ground- and excited-states within a single, symmetry-preserving framework.  Here, symmetry-preserving means that the analysis respects Poincar\'e covariance and satisfies the relevant Ward-Green-Takahashi identities.  Constituent-quark models have hitherto been the most widely applied spectroscopic tools; and whilst their weaknesses are emphasised by critics and acknowledged by proponents, they are of continuing value because there is nothing better that is yet providing a bigger picture.  Nevertheless, they possess no connection with quantum field theory and therefore no connection with QCD; and they are not symmetry-preserving and therefore cannot veraciously connect meson and baryon properties.

A comprehensive approach to QCD will provide a unified explanation of both mesons and baryons.  We have seen that DCSB is a keystone of the Standard Model, evident in the momentum-dependence of the dressed-quark mass function -- Fig.\,\ref{gluoncloud}: it is just as important to baryons as it is to mesons.  The DSEs furnish the only extant framework that can simultaneously and transparently connect meson and baryon observables with this basic feature of QCD, having provided, \emph{e.g}.\ a direct correlation of meson and baryon properties via a single interaction kernel, which preserves QCD's one-loop renormalisation group behaviour and can systematically be improved.  This is evident in Refs.\,\cite{Eichmann:2008ae, Eichmann:2008ef, Eichmann:2009qa, Eichmann:2011ej, Chang:2012cc}.

\begin{figure}[t]
\begin{minipage}[t]{\textwidth}
\begin{minipage}{0.49\textwidth}
\centerline{\includegraphics[clip,width=0.9\textwidth]{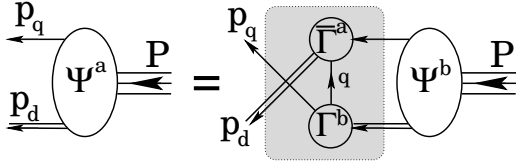}}
\end{minipage}
\begin{minipage}{0.49\textwidth}
\caption{\small
\label{fig:Faddeev} Poincar\'e covariant Faddeev equation.  $\Psi$ is the Faddeev amplitude for a baryon of total momentum $P= p_q + p_d$, where $p_{q,d}$ are, respectively, the momenta of the quark and diquark within the bound-state.  The shaded area is the Faddeev equation kernel: \emph{single line},
dressed-quark propagator; $\Gamma$,  diquark correlation amplitude; and \emph{double line}, diquark propagator.}
\end{minipage}
\end{minipage}
\end{figure}

Let us focus initially on the proton, which is a composite object, whose properties and interactions are determined by its valence-quark content: $u$ + $u$ + $d$, \emph{i.e}.\ two up ($u$) quarks and one down ($d$) quark.  So far as is now known, bound-states seeded by two valence-quarks do not exist; and the only two-body composites are those associated with a valence-quark and -antiquark, \emph{i.e}.\ mesons.  These features are supposed to derive from colour confinement, whose complexities are discussed in Sec.\,\ref{secConfinement}.

Such observations lead one to a position from which the proton may be viewed as a Borromean bound-state \cite{Segovia:2015ufa}, \emph{viz}.\ a system constituted from three bodies, no two of which can combine to produce an independent, asymptotic two-body bound-state.  In QCD the complete picture of the proton is more complicated, owing, in large part, to the loss of particle number conservation in quantum field theory and the concomitant frame- and scale-dependence of any Fock space expansion of the proton's wave function.  Notwithstanding that, the Borromean analogy provides an instructive perspective from which to consider both quantum mechanical models and continuum treatments of the nucleon bound-state problem in QCD.  It poses a crucial question:  \emph{Whence binding between the valence quarks in the proton, \mbox{\rm i.e}.\ what holds the proton together}?

In numerical simulations of lQCD that use static sources to represent the proton's valence-quarks, a ``Y-junction'' flux-tube picture of nucleon structure is produced \cite{Bissey:2006bz, Bissey:2009gw}.  This might be viewed as originating in the three-gluon vertex, which signals the non-Abelian character of QCD and is the source of asymptotic freedom.  Such results and notions would suggest a key role for the three-gluon vertex in nucleon structure \emph{if} they were equally valid in real-world QCD wherein light dynamical quarks are ubiquitous.  However, as we saw in Sec.\,\ref{secConfinement}, they are not; and so a different explanation of binding within the nucleon must be found.

DCSB has numerous corollaries that are crucial in determining the observable features of the Standard Model, some of which are detailed above.  Another particularly important consequence is less well known.  Namely, any interaction capable of creating pseudo-Goldstone modes as bound-states of a light dressed-quark and -antiquark, and reproducing the measured value of their leptonic decay constants, will necessarily also generate strong colour-antitriplet correlations between any two dressed quarks contained within a baryon.  This assertion is based upon evidence gathered in two decades of studying two- and three-body bound-state problems in hadron physics.  No counter examples are known; and the existence of such diquark correlations is also supported by lQCD \cite{Alexandrou:2006cq, Babich:2007ahS}.

The properties of diquark correlations have been charted.  Most importantly, diquarks are confined.  Additionally, owing to properties of charge-conjugation, a diquark with spin-parity $J^P$ may be viewed as a partner to the analogous $J^{-P}$ meson \cite{Cahill:1987qr}.  It follows that scalar, isospin-zero and pseudovector, isospin-one diquark correlations are the strongest in ground-state baryons; and whilst no pole-mass exists, the following mass-scales, which express the strength and range of the correlation and are each bounded below by the partnered meson's mass, may be associated with these diquarks \cite{Cahill:1987qr, Maris:2002yu, Alexandrou:2006cq, Babich:2007ahS}:
$m_{[ud]_{0^+}} \approx 0.7-0.8\,$GeV, $m_{\{uu\}_{1^+}}  \approx 0.9-1.1\,$GeV,
with $m_{\{dd\}_{1^+}}=m_{\{ud\}_{1^+}} = m_{\{uu\}_{1^+}}$ in the isospin symmetric limit.
Realistic diquark correlations are also soft.  They possess an electromagnetic size that is bounded below by that of the analogous mesonic system, \emph{viz}.\ \cite{Maris:2004bp, Roberts:2011wyS}:
$r_{[ud]_{0^+}} \gtrsim r_\pi$, $r_{\{uu\}_{1^+}} \gtrsim r_\rho$,
with $r_{\{uu\}_{1^+}} > r_{[ud]_{0^+}}$.  As with mesons, these scales are set by that associated with DCSB.

%
The interaction depicted in Fig.\,\ref{figInteraction} characterises a realistic class that generates strong attraction between two quarks and thereby produces tight diquark correlations in analyses of the three valence-quark scattering problem.  The existence of such correlations considerably simplifies analyses of baryon bound states because it reduces that task to solving a Poincar\'e covariant Faddeev equation \cite{Cahill:1988dx}, depicted in Fig.\,\ref{fig:Faddeev}.  The three gluon vertex is not explicitly part of the bound-state kernel in this picture of the nucleon.  Instead, one capitalises on the fact that phase-space factors materially enhance two-body interactions over $n\geq 3$-body interactions and exploits the dominant role played by diquark correlations in the two-body subsystems.  Then, whilst an explicit three-body term might affect fine details of baryon structure, the dominant effect of non-Abelian multi-gluon vertices is expressed in the formation of diquark correlations.  Such a nucleon is then a compound system whose properties and interactions are primarily determined by the quark$+$diquark structure evident in Fig.\,\ref{fig:Faddeev}.

\begin{figure}[t]
\begin{minipage}[t]{\textwidth}
\begin{minipage}{0.49\textwidth}
\centerline{\includegraphics[width=0.9\textwidth]{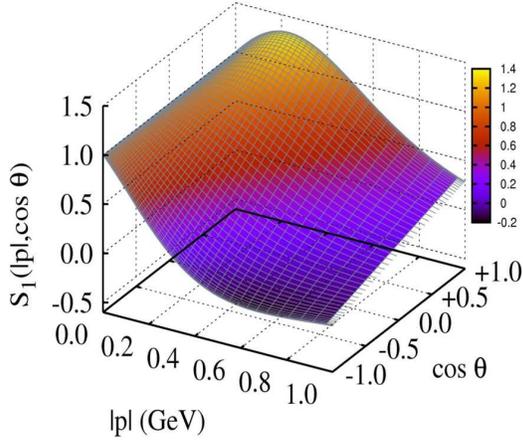}}
\end{minipage}
\begin{minipage}{0.49\textwidth}
\caption{\small
\label{figS1} Representation of the dominant piece in the nucleon's eight-component Poincar\'e-covariant Faddeev amplitude: $s_1(|p|,\cos\theta)$.  In the nucleon rest frame, this term describes that piece of the quark-diquark relative momentum correlation which possesses zero \emph{intrinsic} quark-diquark orbital angular momentum, \emph{i.e}.\ $L=0$ before the propagator lines are reattached to form the Faddeev wave function.  Referring to Fig.\,\ref{fig:Faddeev}, $p= P/3-p_q$ and $\cos\theta = p\cdot P/\sqrt{p^2 P^2}$.  (The amplitude is normalised such that its $U_0$ Chebyshev moment is unity at $|p|=0$.)}
\end{minipage}
\end{minipage}
\end{figure}

The quark$+$diquark structure of the nucleon is elucidated in Fig.\,\ref{figS1}, which provides a representation of the leading component of the nucleon's Faddeev amplitude: with the notation of Ref.\,\cite{Segovia:2014aza}, $s_1(|p|,\cos\theta)$, computed using the Faddeev kernel described therein.  This function describes a piece of the quark$+$scalar-diquark relative momentum correlation.  Notably, in this solution of a realistic Faddeev equation there is strong variation with respect to both arguments.  Support is concentrated in the forward direction, $\cos\theta >0$, so that alignment of $p$ and $P$ is favoured; and the amplitude peaks at $(|p|\simeq M_N/6,\cos\theta=1)$, whereat $p_q \approx P/2 \approx p_d$ and hence the \emph{natural} relative momentum is zero.  In the antiparallel direction, $\cos\theta<0$, support is concentrated at $|p|=0$, \emph{i.e}.\ $p_q \approx P/3$, $p_d \approx 2P/3$.  

A nucleon (and kindred baryons) described by Fig.\,\ref{fig:Faddeev} is a Borromean bound-state, the binding within which has two contributions.  One part is expressed in the formation of tight diquark correlations; but that is augmented by attraction generated by the quark exchange depicted in the shaded area of Fig.\,\ref{fig:Faddeev}.  This exchange ensures that diquark correlations within the nucleon are fully dynamical: no quark holds a special place because each one participates in all diquarks to the fullest extent allowed by its quantum numbers. The continual rearrangement of the quarks guarantees, \emph{inter} \emph{alia}, that the nucleon's dressed-quark wave function complies with Pauli statistics.

One cannot overstate the importance of appreciating that these fully dynamical diquark correlations are vastly different from the static, pointlike ``diquarks'' which featured in early attempts \cite{Lichtenberg:1967zz} to understand the baryon spectrum and to explain the so-called missing resonance problem \cite{Ripani:2002ss, Burkert:2012ee, Kamano:2013iva}.  Modern diquarks are soft and enforce certain distinct interaction patterns for the singly- and doubly-represented valence-quarks within the proton.  On the other hand, the number of states in the spectrum of baryons obtained from the Faddeev equation in Fig.\,\ref{fig:Faddeev} \cite{Chen:2012qrS} is similar to that found in the three-constituent quark model, just as it is in today's lQCD calculations of this spectrum \cite{Edwards:2011jj}.

\section{Roper Resonance}
\begin{figure}[t]
%
%
\centerline{\includegraphics[width=0.8\linewidth]{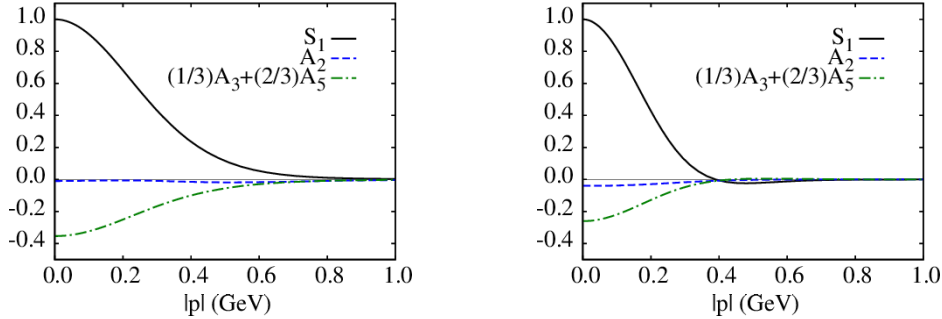}}
\caption{\label{figFA}
\emph{Left}.  Zeroth Chebyshev moment of all $S$-wave components in the nucleon's Faddeev wave function, which is obtained from $\Psi$ in Fig.\,\ref{fig:Faddeev}, by reattaching the dressed-quark and -diquark legs.
\emph{Right}. Kindred functions for the first excited state.
Legend: $S_1$ is associated with the baryon's scalar diquark; the other two curves are associated with the axial-vector diquark; and the normalisation is chosen such that $S_1(0)=1$.}
\end{figure}

 The Roper has long resisted understanding.  JLab experiments \cite{Dugger:2009pn, Aznauryan:2009mx, Aznauryan:2011qj, Mokeev:2015lda} have yielded precise nucleon-Roper ($N\to R$) transition form factors and thereby exposed the first zero seen in any hadron form factor or transition amplitude.  It has also attracted much theoretical attention; but Ref.\,\cite{Segovia:2015hraS} provides the first continuum treatment of this problem using the power of relativistic quantum field theory.  That study begins with a computation of the mass and wave function of the proton and its first radial excitation, using precisely the same framework that was used in a successful unification of nucleon and $\Delta$ properties \cite{Segovia:2014aza}.  The masses are (in GeV):
\begin{equation}
\label{eqMasses}
\mbox{nucleon\,(N)} = 1.18\,,\;
\mbox{nucleon-excited\,(R)} = 1.73\,.
\end{equation}
These values correspond to the locations of the two lowest-magnitude $J^P=1/2^+$ poles in the three-quark scattering problem.  The associated residues are the Faddeev wave functions, which depend upon $(p^2,p\cdot P)$, where $p$ is the quark-diquark relative momentum.  Fig.\,\ref{figFA} depicts the zeroth Chebyshev moment of all $S$-wave components in that wave function.  The appearance of a single zero in $S$-wave components of the Faddeev wave function associated with the first excited state in the three dressed-quark scattering problem indicates that this state is a radial excitation.

It is worth dwelling on the masses in Eq.\,\eqref{eqMasses}.  The empirical values of the pole locations for the first two states in the nucleon channel are \cite{Suzuki:2009njS}: $0.939\,$GeV and $1.36 - i \, 0.091\,$GeV, respectively.  At first glance, these values appear unrelated to those in Eq.\,\eqref{eqMasses}.  However, deeper consideration reveals \cite{Eichmann:2008ae,Eichmann:2008ef} that the kernel in Fig.\,\ref{fig:Faddeev} omits all those resonant contributions which may be associated with the meson-baryon final-state interactions that are resummed in dynamical coupled channels models in order to transform a bare-baryon into the observed state \cite{Suzuki:2009njS,  Kamano:2013iva, Doring:2014qaa}.  This Faddeev equation should therefore be understood as producing the dressed-quark core of the bound-state, not the completely-dressed and hence observable object.

Clothing the nucleon's dressed-quark core by including resonant contributions to the kernel produces a physical nucleon whose mass is $\approx 0.2$\,GeV lower than that of the core \cite{Ishii:1998tw, Hecht:2002ej}.  Similarly, clothing the $\Delta$-baryon's core lowers its mass by $\approx 0.16\,$GeV \cite{Suzuki:2009njS}.   It is therefore no coincidence that (in GeV) $1.18-0.2 = 0.98\approx 0.94$, \emph{i.e}.\ the nucleon mass in Eq.\,\eqref{eqMasses} is 0.2\,GeV greater than the empirical value.  A successful body of work on the baryon spectrum \cite{Chen:2012qr} and nucleon and $\Delta$ elastic and transition form factors \cite{Segovia:2014aza, Roberts:2015deaS} has been built upon precisely this knowledge of the impact of omitting resonant contributions and the magnitude of their effects.

Crucial, therefore, is a comparison between the quark-core mass and the value determined for the mass of the meson-undressed bare-Roper in Ref.\cite{Suzuki:2009njS}, \emph{viz}. (in GeV)
\begin{equation}
\label{eqMassesA}
\begin{array}{l|cc|c}
            & \mbox{R}_{{\rm core}}^{\mbox{\footnotesize \cite{Segovia:2015hraS}}}
            & \mbox{R}_{{\rm core}}^{\mbox{\footnotesize \cite{Wilson:2011aa}}}
            & \mbox{R}_{\rm bare}^{\mbox{\footnotesize \cite{Suzuki:2009njS}}} \\\hline
\mbox{mass} & 1.73 & 1.72 & 1.76
\end{array}\,.
\end{equation}
The bare Roper mass in Ref.\,\cite{Suzuki:2009njS} agrees with both the quark-core result in Eq.\,\eqref{eqMasses} and that obtained using a refined treatment of a vector$\,\otimes\,$vector contact-interaction \cite{Wilson:2011aa}.  This is notable because all these calculations are independent, with just one common feature; namely, an appreciation that measured hadrons can realistically be built from a dressed-quark core plus a meson-cloud.

The agreement in Eq.\,\eqref{eqMassesA} is suggestive but not conclusive.  As noted above, precise empirical information is available on the nucleon-Roper transition form factors.  Thus, if the picture described herein is valid, then combining the solutions of the Faddeev equation in Fig.\ref{fig:Faddeev} for both the ground-state nucleon and its radial excitation should produce transition form factors that possess an understandable connection with available data and, indeed, match in accuracy the predictions for the nucleon and $\Delta$-baryon elastic and transition form factors obtained using the same approach \cite{Segovia:2014aza, Roberts:2015deaS}.

The QCD-based Faddeev equation predicts the existence of diquark correlations within baryons; and it is interesting to compare the diquark content of the nucleon and its radial excitation.  That information is contained in the zero-momentum value of the elastic Dirac form factor \cite{Wilson:2011aa, Segovia:2014aza}:
\begin{equation}
\label{Pdiquark}
\begin{array}{l|cc}
        & N    & R    \\\hline
P_{J=0} & 62\% & 62\%  \\
P_{J=1} & 38\% & 38\%\\
\end{array}\,;
\end{equation}
namely, the relative strength of scalar and axial-vector diquark correlations in the nucleon and its radial excitation is the same. 

\begin{figure}[t]
\begin{minipage}[t]{\textwidth}
\begin{minipage}{0.49\textwidth}
\centerline{\includegraphics[clip,width=0.85\linewidth]{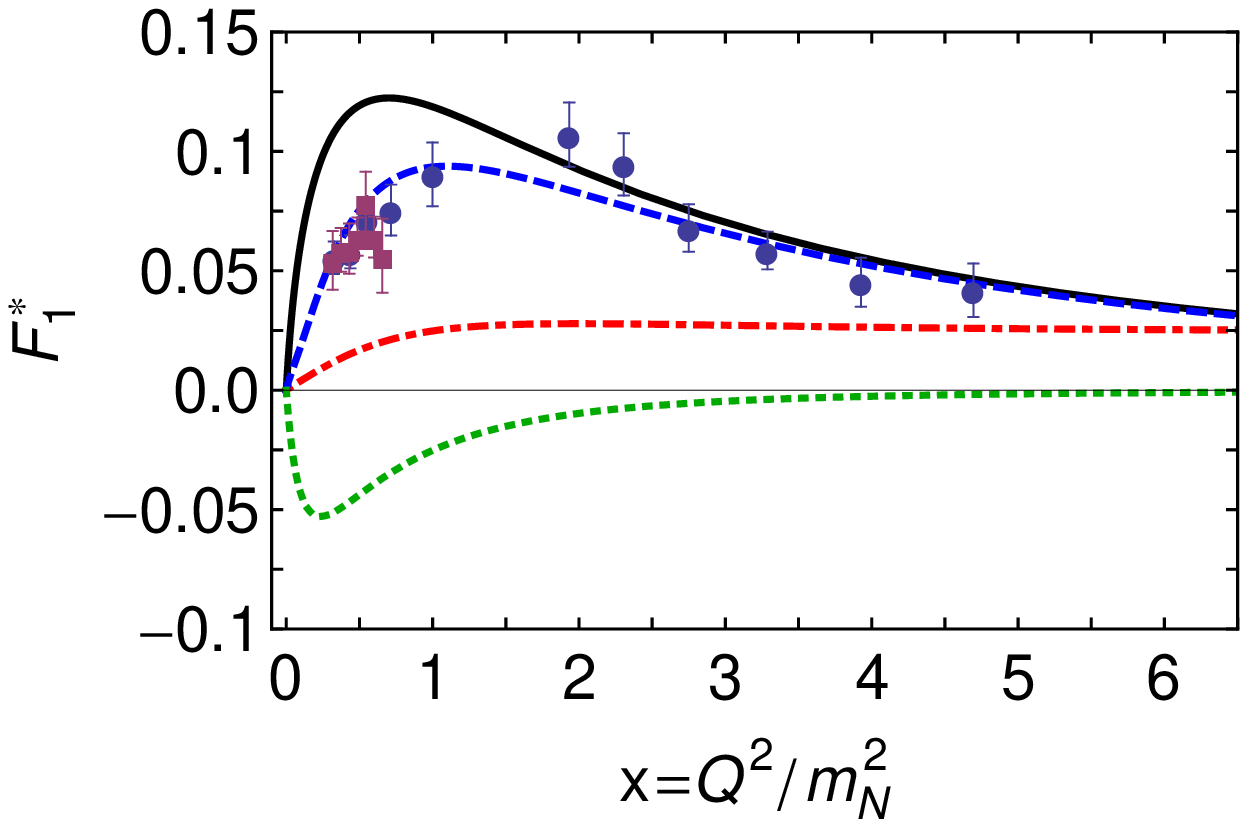}}
\end{minipage}
\begin{minipage}{0.49\textwidth}
\centerline{\includegraphics[clip,width=0.85\linewidth]{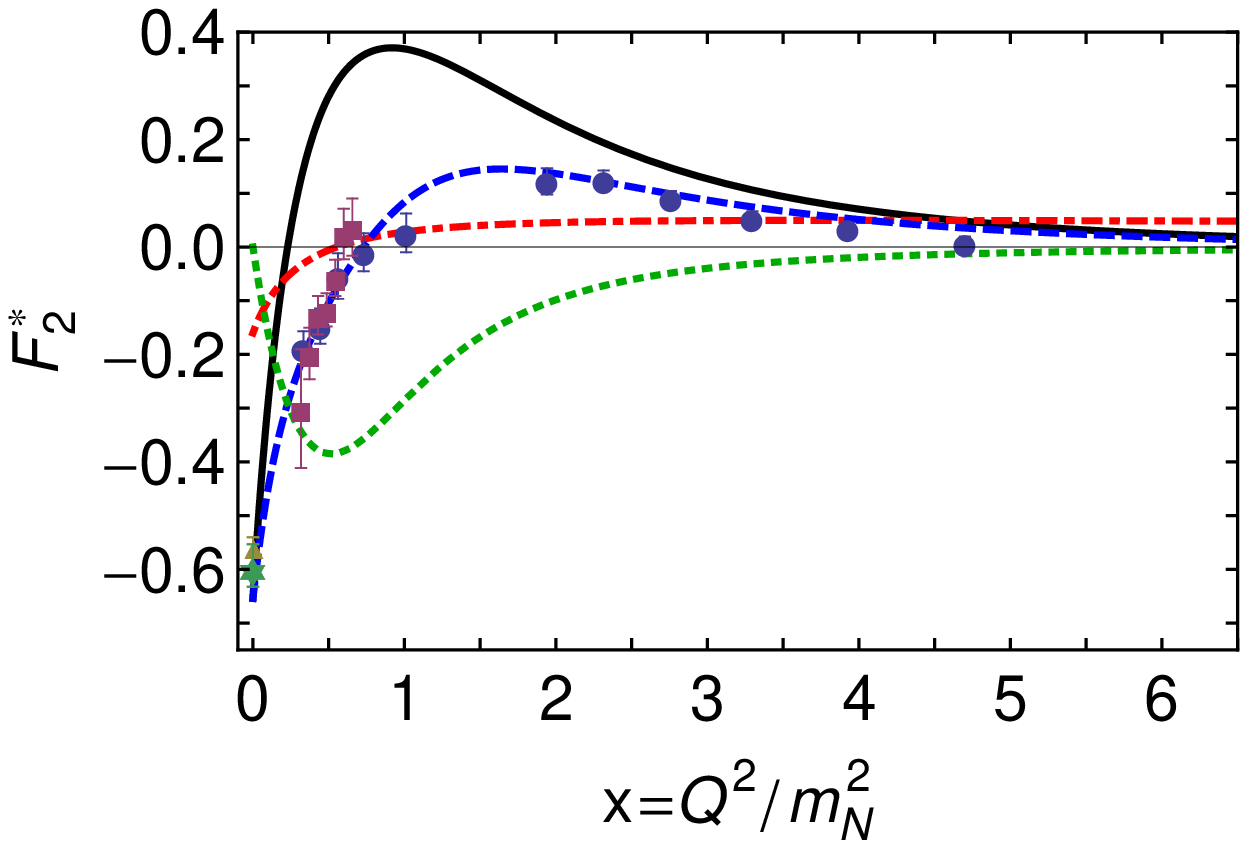}}
\end{minipage}
\end{minipage}
\caption{\label{figFT}
\emph{Left} -- Dirac transition form factor, $F_{1}^{\ast}(x)$, $x=Q^2/m_N^2$.  Solid (black) curve, our prediction; dot-dashed (red) curve, contact-interaction result \cite{Wilson:2011aa}; dotted (green) curve, inferred meson-cloud contribution; and dashed (blue) curve, anticipated complete result.
\emph{Right} -- Pauli transition form factor, $F_{2}^{\ast}(x)$, with same legend.
Data in both panels: circles (blue) \cite{Aznauryan:2009mx};
triangle (gold) \cite{Dugger:2009pn};
squares (purple) \cite{Mokeev:2012vsa};
and star (green) \cite{Agashe:2014kda}.}
\end{figure}

The transition form factors are displayed in Fig.\,\ref{figFT}.  The DSE predictions agree quantitatively in magnitude and qualitatively in trend with the data on $x\gtrsim 2$.  Nothing was tuned to achieve those results.  Instead, the nature of the DSE prediction owes fundamentally to the QCD-derived momentum-dependence of the propagators and vertices employed in formulating the problems.  This point is further highlighted by the contact-interaction result: momentum-independent propagators and vertices yield predictions that disagree both quantitatively and qualitatively with the data.  Experiment is evidently a sensitive tool with which to chart the nature of the quark-quark interaction and hence discriminate between competing theoretical hypotheses; and it is plainly settling upon an interaction that produces the momentum-dependent dressed-quark mass which characterises QCD \cite{Bowman:2005vx, Bhagwat:2006tu, Roberts:2007ji}.  
The mismatch between the DSE predictions and data on $x\lesssim 2$ is also revealing.  Meson-cloud contributions are expected to be important on this domain \cite{Segovia:2014aza, Roberts:2015deaS}.  An inferred form of that contribution is provided by the dotted (green) curves in Fig.\,\ref{figFT}.
These curves have fallen to just 20\% of their maximum value by $x=2$ and vanish rapidly thereafter so that the DSE predictions alone remain as the explanation of the data.  Importantly, the existence of a zero in $F_{2}^{\ast}$ is not influenced by meson-cloud effects, although its precise location is.  (The same is true of the $p\to\Delta^+$ electric transition form factor.)  Thus any realistic approach to the $p\to R$ transition must describe a zero in $F_{2}^{\ast}$.

Numerous properties of the dressed-quark core of the proton's radial excitation were computed in Ref.\cite{Segovia:2015hraS}.  In all cases they provide an excellent understanding of data on the proton-Roper transition and related quantities derived using coupled channels models.  The DSE analysis is based on a sophisticated framework for the three-quark bound-state problem; all elements employed possess an unambiguous link with analogous quantities in QCD; and no parameters were varied in order to achieve success.  One may thus conclude that the Roper resonance is at heart the nucleon's first radial excitation, consisting of a dressed-quark core augmented by a meson cloud that reduces its (Breit-Wigner) mass by approximately 20\%.  The analysis shows that a meson-cloud obscures the quark core from long-wavelength probes; but that it is revealed to probes with $Q^2 \gtrsim 3 m_N^2$.  This feature is typical of nucleon-resonance transitions; and hence measurements of resonance electroproduction on this domain can serve as an incisive probe of quark-gluon dynamics within the Standard Model, assisting greatly in mapping the evolution between the nonperturbative and perturbative domains of QCD.

\section{Summary}
It is worth reiterating a few significant points.
Owing to the conformal anomaly, both gluons and quarks acquire mass dynamically in QCD.  Those masses are momentum dependent, with large values at infrared momenta.
The appearance of these nonperturbative running masses is intimately connected with confinement and DCSB; and the relationship between those phenomena entails that in a Universe with light-quarks, confinement is a dynamical phenomenon.  Consequently, flux tubes are not the correct paradigm for confinement and it is meaningless to speak of linear potentials and Regge trajectories.
In exploring the connection between QCD's gauge and matter sectors, top-down and bottom-up DSE analyses have converged on the form of the renormalisation-group-invariant interaction in QCD.  This outcome paves the way to parameter-free predictions of hadron properties.
Decades of studying the three valence-body problem in QCD have provided the evidence necessary to conclude that diquark correlations are a reality; but diquarks are complex objects so that their existence does not restrict the number of baryon states in any obvious way.  This effort has led us to a sophisticated understanding of the nucleon, $\Delta$-baryon and Roper resonance: all  may be viewed as Borromean bound-states, and the Roper is at heart the nucleon's first radial excitation.
The progress summarised herein highlights the capacity of DSEs in QCD to connect the quark-quark interaction, expressed, for instance, in the dressed-quark mass function, $M(p^2)$, with predictions for a wide range of hadron observables; and therefore serves as strong motivation for new experimental studies of nucleon elastic and transition form factors, which exploit the full capacity of JLab\,12 in order to chart $M(p^2)$ and thereby explain the origin of more than 98\% of the visible mass in the Universe.

\medskip

%
\hspace*{-\parindent}\textbf{Acknowlegments}. The material described in this contribution is drawn from work completed in collaboration with numerous excellent people, to all of whom I am greatly indebted.
I would also like to thank R.~Gothe, T.-S.\,H.~Lee, V.~Mokeev and T.~Sato for constructive input;
and to express my gratitude to the sponsors of \emph{The 10th International Workshop on the Physics of Excited Nucleons (NSTAR2015)}, whose support helped enable my participation; and to the organisers and my hosts, who ensured both that the meeting was a success and my participation was enjoyable and fruitful.
Work supported by  U.S.\ Department of Energy, Office of Science, Office of Nuclear Physics, under contract no.~DE-AC02-06CH11357.

\providecommand{\newblock}{}

\end{document}